\documentclass[aps,pre,twocolumn,showpacs,groupedaddress]{revtex4}

\usepackage{amsmath}
\usepackage{amsfonts}
\usepackage{amssymb}
\usepackage{bm}
\usepackage{color}
\usepackage{graphicx}

\begin{document}
\title{Statistics of trajectories in two-state master equations}
\author{Andrew D. Jackson and Simone Pigolotti}
\affiliation{The Niels Bohr International Academy, The Niels Bohr Institute, 
Blegdamsvej 17, DK-2100 Copenhagen, Denmark}

\begin{abstract}
We derive a simple expression for the probability of trajectories of a
master equation. The expression is particularly useful when the number
of states is small and permits the calculation of observables that 
can be defined as functionals of whole trajectories.  We illustrate the 
method with a two-state master equation, for which we calculate the 
distribution of the time spent in one state and the distribution of 
the number of transitions, each in a given time interval. These two 
expressions are obtained analytically in terms of modified Bessel functions.
\end{abstract}
\maketitle

The evolution of many systems in physics, chemistry and biology is
properly described by master equations.  This description is adequate
when the system under consideration has discrete states and when the
rate of jumping from one state to another does not depend on the
history of the system, i.e. when the Markov property holds. In recent
years, this description has been successfully applied to a plethora of
new problems in several fields. As examples, master equations are
commonly used in biochemistry to understand the fluctuations of
chemical concentrations inside the cell \cite{elowitz}. In statistical
physics, they can provide a simple description of non-equilibrium
systems, useful for testing the validity of fluctuation relations
\cite{lebowitz,esposito,harris}.

From a technical point of view, master equations now constitute a 
well established field of research, and many techniques have been
developed which permit their analytical or numerical treatment
\cite{gardiner,gillespie,pigo}. In complicated cases, these techniques
permit calculation of the steady-state probabilities, $P_n$, of being 
in state $n$. In simpler cases, it is sometimes possible to solve
equations in time in order to determine the propagator, $p(n,t|n_0,0)$, 
that gives the probability of being in state $n$ at time $T$ starting
from a state $n_0$ at time $t=0$.

For many practical purposes, determination of the propagator is
sufficient, since many interesting observables can be expressed as a
function of the propagator. There are, however, observables that
cannot be obtained conveniently from the propagator, including in
particular quantities which are more easily expressed as {\em
  functionals\/} of entire trajectories. Examples include the
distribution of the time spent in a given state and the probability of
observing a given number of transitions, both for a fixed time
interval.  Functional methods are well known for continuous stochastic
process, where techniques have been developed in parallel to those
used in quantum mechanics \cite{itch}. There are fewer examples of
functional methods for discrete processes \cite{peliti,cardy}.  These
methods are often field theoretic in nature and involve complications
such as renormalization which one would like to avoid in simple
discrete systems.

In this paper, we present a simple way to calculate probabilities of
the trajectories of master equations. The method is straightforward,
rigorous and does not require any specific assumptions on the
equation. It is particularly useful when the number of states
available to the system is small, where it is possible to obtain
closed analytical expression for several interesting observables.  We
study as example of our method general two-state systems that, despite
their simplicity, have many non-trivial applications in problems
related to single-molecule spectroscopy (see,
e.g. \cite{margolin,shikerman}) and biophysics (see, e.g.,
\cite{bonnet,zwanzig, ritort}).  Specifically, we calculate the
probability of observing a given number of transitions, $N$, in a
time, $T$, and the distribution of time spent in one of the two states
in a time $T$.  Each of these quantities can be expressed in terms of
modified Bessel functions.

We consider a master equation:

\begin{equation}\label{mastereq}
\frac{d}{dt}P_n= \sum_m W_{mn}P_m-W_{nm}P_n,
\end{equation}
where $P_n(t)$ is the time-dependent probability of being state $n$ 
and $W_{mn}$ is the transition rate from state $m$ to $n$. For 
convenience we also define:
\begin{equation}
W^{out}_n = \sum_i W_{ni},
\end{equation}
the total out-rate of state $n$.  The probability that, in a time 
$T$, the trajectory visits a pre-determined sequence of states $n_0, \, 
n_1, \, n_2 \, \dots \, n_N$ then becomes
\begin{eqnarray}\label{trajprob}
\mathcal{P}(n_0,n_1,\dots n_N;T)=\int_0^T dt_1 \int_{t_1}^T dt_2
\dots \int_{t_{n-1}}^T dt_N 
\times\nonumber\\\times {\rm e}^{-W^{out}_0 t_1} W_{n_0n_1}
e^{-W^{out}_1 (t_2-t_1)} \dots W_{n_{N-1}n_N}{\rm e}^{-W^{out}_{n} (T-t_n)}.
\end{eqnarray}
\begin{figure}[htb]
\includegraphics[width=8.5cm]{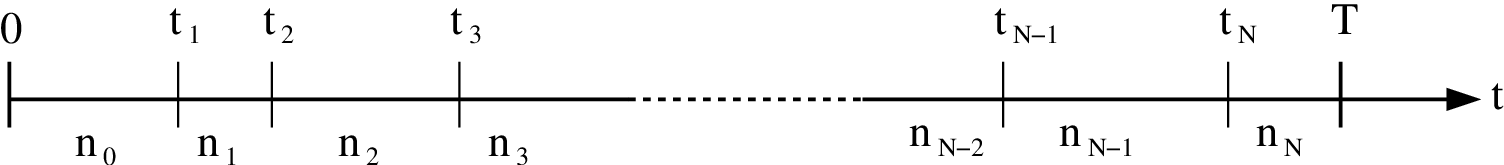}
\caption{Trajectory of a master equation as a function of time. The
integral in eq.\,(\ref{trajprob}) is over the times of the $N$ transition 
points.}
\end{figure}
This expression can be understood by noticing that the integrand
represents the probability density in time of the $N$ consecutive
transitions according to the master equation (see Fig. 1). By summing
over all trajectories having pre-determined properties, one can
reconstruct the full statistics of the stochastic process. An obvious
example is the propagator, that can be evaluated as the sum over all
trajectories that start in a given state, $n_0$, at time $t=0$ and end
in a state $n_f$ at time $T$:
\begin{equation}
p(n_f,T|n_0,0)=\sum\limits_{N=0}^{\infty}\sum\limits_{\{n_1 \dots n_{N-1}\}} 
\mathcal{P}(n_0,n_1,\dots n_N=n_f;T) \, .
\end{equation}
Note that all probabilities are properly normalized; in particular,
the propagator satisfies the closure relation
$\sum_{n_F}p(n_f,T|n_0,0)=1$.

The above expressions become particularly useful when the number of
distinct states visited by the system is small. In this case, it is
convenient to rearrange the integrals in eq.\,(\ref{trajprob})
by grouping together all time intervals in which the system is in the
same state. If $k_i$ is the number of times the system visits state
$i$ on a given trajectory, one finds
\begin{eqnarray}\label{traj_prob_states}
\mathcal{P}(n_0,n_1,n_2\dots n_N;T)= W_{n_0n_1}W_{n_1n_2}\dots 
W_{n_{N-1}n_N}\times\nonumber\\
\times\int_0^T \delta(\sum_i t_i-T)\prod_i 
\left[\exp\left(-W^{out}_i t_i\right)\frac{t_i^{k_i-1}}{(k_i-1)!}\right] dt_i\ 
\end{eqnarray}
where the index $i$ runs over all states visited by the system at
least once in the given sequence.

As an example, we consider the simple case of a master equation with
two states, $+$ and $-$, with transition rates $k_+$ (from $-$ to $+$)
and $k_-$ (from $+$ to $-$). In spite of its simplicity, this case is
of interest for many physical and biological problems
\cite{margolin,shikerman,bonnet,zwanzig,ritort}.  We will show that
eq.\,(\ref{traj_prob_states}) allows analytic calculation of the
probabilities of different classes of trajectories.  This makes it
possible, for example, to obtain closed expressions for the
probability of observing a given number of transitions in a time $T$
and for the probability of spending a given time in states $\pm$
during a time $T$. For convenience, we introduce here the total rate
$k_T = k_{+}+k_{-}$ and the equilibrium probabilities,
$P^{eq}_{+}=k_{+}/k_T$ and $P^{eq}_{-}=k_{-}/k_T$. In this case, we
can immediately write the probabilities of all possible trajectories
according to eq.\,(\ref{traj_prob_states}). The simplest trajectories
are evidently those in which there is no transition in the interval

$[0,T]$:
\begin{eqnarray}\label{prob_notr_twostates}
\mathcal{P}(+;T)={\rm e}^{-k_-T}\nonumber\\
\mathcal{P}(-;T)={\rm e}^{-k_+T}.
\end{eqnarray}
The determination of general trajectories is simplified by having 
only two states, since trajectories can only alternate between them.  
It is then convenient to classify trajectories according to: a) the 
initial state ($+$ or $-$), b) the total time $T$, c) the total time 
spent in state $\pm$, $t_{\pm}$, and d) the total number of transitions, 
$N$.  This is sufficient to characterize a general term in 
eq.\,(\ref{traj_prob_states}).  Note that slightly different 
expressions are obtained for $N$ even and for $N$ odd. The 
result is:
\begin{eqnarray}\label{probab_twostates}
\mathcal{P}(-,T,t_+,N_{even})&=&\frac{[k_+(T-t_+)]^{\frac{N}{2}}
(k_-t_+)^{(\frac{N}{2}-1)}}{\frac{N}{2}!(\frac{N}{2}-1)!}\ 
k_-\ {\rm e}^{-r}\nonumber\\
\mathcal{P}(-,T,t_+,N_{odd})&=&\frac{[k_+(T-t_+)(k_-t_+)]^{\frac{N-1}{2}}}
{\frac{N-1}{2}!\frac{N-1}{2}!}\ k_+\  {\rm e}^{-r}\nonumber\\
\mathcal{P}(+,T,t_+,N_{even})&=&\frac{[k_+(T-t_+)]^{\frac{N}{2}-1}
(k_-t_+)^{\frac{N}{2}}}{\frac{N}{2}!(\frac{N}{2}-1)!}\ k_+ \ {\rm e}^{-r} \nonumber\\
\mathcal{P}(+,T,t_+,N_{odd})&=&\frac{[k_+(T-t_+)k_-t_+]^{\frac{N-1}{2}}}
{\frac{N-1}{2}!\frac{N-1}{2}!}\ k_-\ {\rm e}^{-r}.
\end{eqnarray}
where $r=[k_-t_++k_+(T-t_+)]$. These equations describe all trajectories
with $N > 0$ while eqn.\,(\ref{prob_notr_twostates}) describes the two
trajectories with $N=0$.  Note, however, that
eqn.\,(\ref{prob_notr_twostates}) describes probabilities while
eqns.\,(\ref{probab_twostates}) are probability densities in $t_+$. To
obtain consistent notation, the two expressions in
eq.\,(\ref{prob_notr_twostates}) should be multiplied by
$\delta(t_{+}-T)$ and $\delta(t_{+})$, respectively.  This formalism
allows us to calculate the distribution of time spent in a state
during a time interval $T$, $g(t_{\pm}|T)$.  Drawing the initial state
from the equilibrium distribution $(P^{eq}_{+},P^{eq}_{-})$, we find
\begin{equation}
g(t_{+}|T)=P^{eq}_{+} \, \sum\limits_{N=0}^{\infty} \, \mathcal{P}(+,t_+,N) 
+ P^{eq}_-\sum\limits_{N=0}^{\infty} \mathcal{P}(-,T,t_+,N).
\end{equation}
Inserting eqns.\,(\ref{prob_notr_twostates}) and (\ref{probab_twostates})
into this expression and summing the series, we obtain 
\begin{eqnarray}\label{fung}
&g(t_{+}|T)=P^{eq}_{-} \, {\rm e}^{-k_{+}T}\delta(t_{+}) + P^{eq}_{+} \, 
{\rm e}^{-k_{+}T}\delta(T-t_+)+\nonumber\\
&+ {\rm e}^{-r}\left[\left(\frac{k_-}{t_{+}} 
+ \frac{k_{+}}{(T-t_{+})}\right)\frac{z}{k_T} I_1 (2z)  + 
\frac{2k_{+} k_{-}}{k_T}I_0(2z)\right] \ ,
\end{eqnarray}
where $r$ is same as in eq. (\ref{probab_twostates}),
$z=\sqrt{k_{-}t_{+}k_{+}(T-t_{+})}$ and $I_0(z)$ and $I_1(x)$ are
modified Bessel functions. Notice that this result can be obtained in
a less direct way by means of the Anderson formalism
\cite{berezhkovskii}. Note also that the propagators can be obtained
by an integration over $t_{+}$ of the various terms contributing to
$g( t_{\pm} | T)$.
\begin{figure}[htb]
\includegraphics[width=8.5cm]{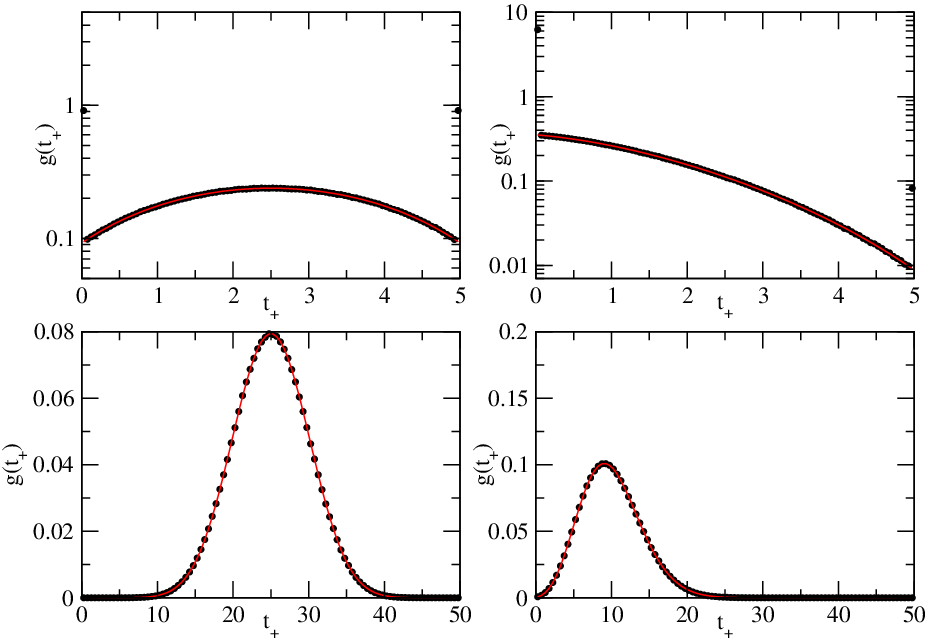}
\caption{A comparison of a simulation and eq.\,(\ref{fung}) for 
the function $g(t_{+}|T)$.  The parameters are $T=5$ (top figures) and $T=50$
(bottom figures). The rates are $k_{+}=k_{-}=0.5$ (left figures) and $k_{-}=0.8$ 
and $k_{+}=0.2$ (right figures). Lines (red on-line) are the analytic curves; 
the black point are averages over $10^7$ simulations of the master
equation. Notice the effect of the Delta functions (first and last
point) in the top figures, where the probabilities are in log scale.
(We do not plot the delta functions in the analytic curves).}\label{figgn}
\end{figure}

In Fig.\,(\ref{figgn}) we plot the function $g(t_{+}|T)$ for several
values of the parameters, and we compare it with simulations of the
master equations. In all cases studied, there is perfect agreement 
between the simulations and the present analytic result.

An interesting limit of eq.\,(\ref{fung}) is that of large $T$. Using 
the asymptotic expression $\lim_{x \to \infty} \left[ I_\nu(x) \right] 
= {\rm exp}\left( x/\sqrt{2\pi z} \right)$, we see that the leading term 
in $1/T$ is
\begin{equation}
g(t_{+}|T) \approx \sqrt{\frac{2}{\pi z}}\frac{k_{+}k_{-}}{k_T} \, 
{\rm e}^{-\left( \sqrt{(k_{-}t_{+})}-\sqrt{k_{+}(T-t_{+})} \right)^2} \ ,
\end{equation}
which has exponential tails expected from large deviation arguments 
\cite{touchette}.

Another issue that can be addressed in this framework is the
probability, $h(N)$, of observing precisely $N$ transitions 
in a time interval of $T$. This is simply:
\begin{equation}
h(N)=\int_0^T dt_+ [P^{eq}_{-} \, \mathcal{P}(-,T,t_+,N) +  
P^{eq}_{+}\mathcal{P}(+,T,t_+,N)].
\end{equation}
Using the above expressions for the various terms and performing the 
integral, we find two different expressions, one for $N$ odd:
\begin{eqnarray}\label{odddistr}
h(N)=\frac{2\sqrt{\pi}(k_{+}k_{-})^{(N+1)/2}}{((N-1)/2)!k_T} \left( 
\frac{T}{k_{-}-k_{+}}\right)^{\frac{N}{2}}\times\nonumber\\\times {\rm e}^{-(k_{+}+k_{-})T/2} \, 
I_{N/2}\left( \zeta \right)
\end{eqnarray}
and one for $N$ even:
\begin{eqnarray}\label{evendistr}
&h(N)=\frac{\sqrt{\pi}T (k_{-}k_{+})^{(N/2)}}{2 k_T (N/2)!} \,
\left( \frac{T}{k_{-}-k_{+}} \right)^{\frac{(N-1)}{2}}
\!\!\!{\rm e}^{-(k_{+}+k_{-})\frac{T}{2}} \times\nonumber\\
&\times \left[(k_{-}+k_{+}) \, I_{(N-1)/2} \left( \zeta  \right)
+ (k_{-}-k_{+}) \, I_{(N+1)/2} \left( \zeta \right) \right]\qquad 
\end{eqnarray}
with $\zeta = (k_{-}-k_{+} )T/2$.

\begin{figure}[htb]
\includegraphics[width=8.5cm]{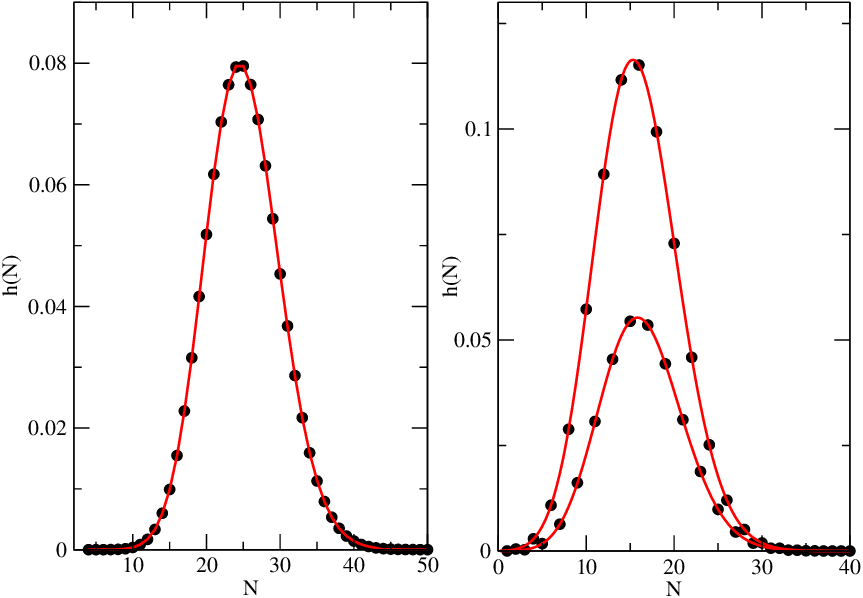}
\caption{The probabilities, $h(N)$, of observing $N$ transitions in a
time $T=50$. Transition rates are (left) $k_{+}=k_{-}=0.5$ (right) and
$k_{+}=0.2$ and $k_{-}=0.8$ (left).  The points represent statistics
collected over $10^7$ simulations.  The solid lines (red on-line) are
the analytic results of eqns.\,(\ref{odddistr}) and
(\ref{evendistr}). The left figure corresponds to the symmetric limit
with $k+{+} = k_{-} = 0.5$ in which both distributions collapse into a
Poisson distribution. Notice the even-odd asymmetry in the right
figure.}\label{fighn}
\end{figure}

In evaluating the two expressions above, care must be taken to pick up
the proper branch of the half-integer powers according to the
requirement that the function $h(N)$ should be real and positive.  In
Fig.\,(\ref{fighn}) we compare the distribution $h(N)$ with
simulations of the master equation.  Here, too, perfect agreement is
found.  The left panel shows a symmetric case with $k_{+}=k_{-}=0.5$
for which eqns.\,(\ref{odddistr}) and (\ref{evendistr}) each have as
limit a Poisson distribution, $h(N)=\lambda^N {\rm e}^-{\lambda}/N!$
with $\lambda=T/k_{+}=T/k_{-}$. The right panel, for the case
$k_+=0.2$ and $k_-=0.8$, is less trivial. The asymmetry in the rates
is reflected in a difference between the distributions for $N$ even
and $N$ odd.  This corresponds to the physical fact that one of the
states is short-lived and the other long-lived, so that is more likely
to observe an even number of transitions. In the asymmetric case,
accurate numerical studies indicate that the average number of
transitions is $\bar{N}=T/\tilde{k}$ with $\tilde{k}= k_{+} k_{-}/
[2(k_{+}+k_{-})]$.

In summary, we have shown that the probability distributions
associated with the trajectories of master equations can be expressed
in general as a product over single-state properties.  This can be
particularly useful for systems composed of a few states as
demonstrated by the exact determination of several statistical
quantities of two-state master equations for which results can be
expressed simply in terms of modified Bessel functions.  While the
methods presented here can be applied to the evaluation of individual
trajectories in more complex problems, summation over all trajectories
becomes increasingly difficult as the number of states increases. If,
however, almost all rates are small, the dynamics of the system can be
dominated by a relatively small number of trajectories. For example,
this is often the case in chemical kinetics, where average reaction
paths may be well defined even for high-dimensional dynamics
\cite{zuckermann}.  In such cases, our methods could provide a way to
detect these dominant trajectories and to assess their probabilities.

\begin{acknowledgments}
  We would like to thank E. Barkai for pointing out relevant
  references.  S. P. wishes to thank J. Ferkinghoff-Borg, J. Fonslet,
  M. H. Jensen and S. Krishna for stimulating discussion.
\end{acknowledgments}

\end{document}